\documentclass[superscriptaddress,amsmath,amssymb,10pt,pra,floatfix,twocolumn]{revtex4-2}
\usepackage{graphicx}
\usepackage{color}
\usepackage{hyperref}
\usepackage{verbatim}
\usepackage{xfrac}
\usepackage{braket}
\usepackage{amsmath}
\newcommand{\revision}[1]{\textcolor{black}{#1}} 

\begin{document}
	\title{Vortex phases and domain walls in trapped spinor Bose-Einstein condensates with inhomogeneous spin-orbital-angular-momentum coupling}
	\author{O. O. Prykhodko}
\address{Department of Physics, Taras Shevchenko National University of Kyiv, Volodymyrska Str. 64/13, Kyiv 01601, Ukraine}
\author{L. V. Zadorozhna}
\address{Department of Physics, Taras Shevchenko National University of Kyiv, Volodymyrska Str. 64/13, Kyiv 01601, Ukraine}
\address{Niels Bohr Institute, Jagtvej 155A, 2200, K{\o}benhavn, Denmark}
	
\begin{abstract}
	
We investigate the ground-state structures and vortex configurations in a two-component Bose-Einstein condensate (BEC) under the influence of spin-orbital-angular-momentum coupling (SOAMC) with a high spatial inhomogeneity and high characteristic orbital angular momentum. By modulating the coupling strength, we uncover two distinct quantum phases: a stripe phase at low coupling strengths and a new vortex-necklace phase at higher coupling intensities. The latter is characterized by vortices forming a ring-shaped structure that acts as a domain wall, a unique phase boundary between a central stripe phase and an outer single-momentum phase. For a better understanding of this new mixed phase of the system, we develop an analytical model to describe the domain wall radius as a function of coupling strength, which aligns well with numerical simulations. Our findings contribute to the understanding of SOAMC-driven quantum phase transitions and domain wall formation, offering new insights into topological phenomena in ultracold atomic systems. 

\end{abstract}

	\maketitle

\section{Introduction}

Implementation of synthetic spin-orbit coupling (SOC) in ultracold atomic systems has enabled the investigation of numerous quantum phenomena within a highly controllable environment. In particular, SOC has enabled the study of exotic structures in atomic Bose-Einstein condensates (BECs), where phenomena such as vortex formation, topological defects, and various quantum phase transitions provide critical insights into the interplay between spin-orbit coupling and nonlinear atomic interactions \cite{Goldman_2014,Zhai2015,ZHANG201975}. Early implementations of synthetic SOC in both bosonic \cite{Lin2011} and fermionic \cite{PhysRevLett.109.095301} quantum gases employed two-photon Raman coupling to link atomic hyperfine states and transfer a part of the photon momentum to the atom. This technique, together with several other approaches \cite{luo2016tunable,PhysRevLett.117.185301}, has laid the groundwork for more complex forms of SOC beyond the linear-momentum transfer \cite{Huang2016,Wu2016,PhysRevLett.118.190401}. One prominent example is spin-orbital-angular-momentum coupling (SOAMC), which incorporates the transfer of orbital angular momentum (OAM) using ``twisted'' light modes, e.g., Laguerre-Gaussian laser beams \cite{PhysRevLett.121.113204,PhysRevLett.122.110402,PhysRevA.93.013629,PhysRevA.94.033627,PhysRevA.91.063627,PhysRevResearch.2.033152,Chiu_2020}.

SOAMC has attracted attention due to its ability to induce a variety of quantum phases within BECs, including intricate topological states  and multi-vortex configurations. 
Recent research has explored SOC-driven phases and phase transitions in harmonically trapped \cite{PhysRevA.91.033630,PhysRevA.102.063328,liu2024vortex} or toroidal condensates \cite{PhysRevA.91.063627,PhysRevA.105.023320}, highlighting important effects of the condensate geometry on the phase diagram. 
Despite tremendous progress, certain aspects of SOAMC remain less understood, particularly regarding high-OAM configurations, which also means strong spatial inhomogeneity of Raman beams. Specifically, the effect of spatially varying Raman coupling and condensate density on the quantum phase landscape remains as an open question.

In this work, we revisit a previously studied SOAMC system in a harmonic trapping potential, with a focus on the impact of high OAM carried by the Raman beams and associated spatial inhomogeneity of the coupling strength on the condensate’s phase structure. 
This inhomogeneity leads a stark difference between red- and blue-detuned variants of the Raman coupling. It also may lead to the formation of a vortex "necklace" that acts as a domain wall, marking a boundary between spatially separated phases within the condensate. Recent studies have shown that such vortex structures can act as phase boundaries and support unique dynamical behaviors within quantum gases \cite{Yao2022,PhysRevResearch.5.023109,PhysRevA.70.043624}. In this work, we propose a relatively simple analytical model for describing such vortex-necklace states. By systematically analyzing these domain wall structures, we aim to expand the understanding of SOAMC-driven quantum phase transitions and provide a framework for interpreting mixed-phase configurations in spinor BECs. 

\section{Model} 

We consider an ultracold gas of bosonic atoms with several
ground state sublevels, for instance, the $F = 1$ states of $^{23}$Na.
The atoms are 
subjected to a static magnetic field along $z$, lifting the Zeeman
degeneracy and providing the quantization axis. The atoms are further illuminated by two laser beams
copropagating along $z$ with opposite circular
polarizations. 
These two laser beams are far-detuned 
from a single-photon atomic transition and create a two-photon resonant Raman
coupling between the Zeeman sublevels $m_F = \pm 1$ with a $\Lambda$ scheme (see Fig. \ref{fig:level-scheme}).
In the basis of atomic states $\{\ket{e},\ket{-1},\ket{1}\}$ and under the rotating-wave approximation, the atom-light Hamiltonian is given by
\begin{figure}[tbp]
	\centering
	\includegraphics[width=0.5\linewidth]{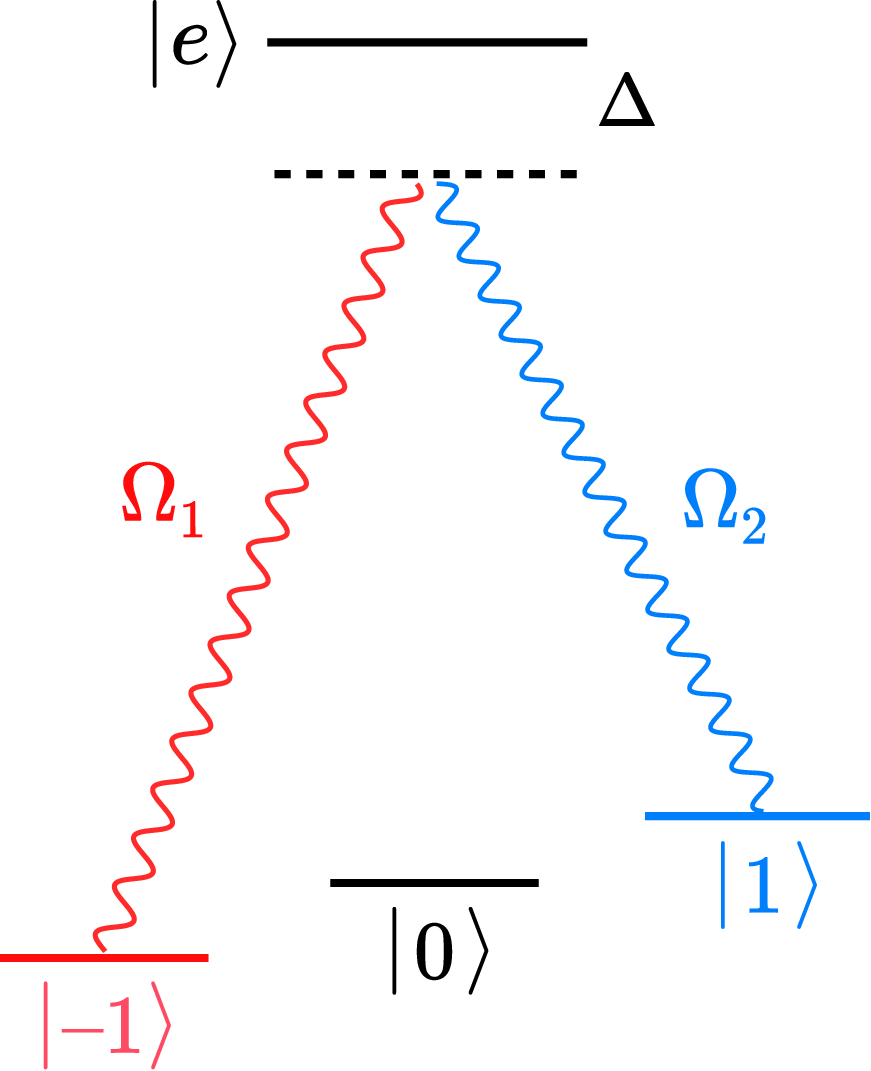}
	\caption{A lambda scheme: The electron is transferred between the hyperfine substates $\ket{F=1,m_F=-1}$ and $\ket{F=1,m_F=1}$ via the two-photon process.}
	\label{fig:sp_pd}
\end{figure}
\begin{equation}\label{fig:level-scheme}
	\mathcal{H}_{\Lambda} = \frac{\hbar}{2} \begin{pmatrix}
		 2 \Delta & \Omega_1^* & \Omega_2^*\\ 
		 \Omega_1 & 0 & 0\\ 
		 \Omega_2 & 0 & 0
	\end{pmatrix},
\end{equation}
\revision{where $\Omega_1$ and $\Omega_2$ are the Raman coupling strengths, $\Delta = \omega_1 - \omega_1^{(\mathrm{L})} = \omega_2 - \omega_2^{(\mathrm{L})}$ is the single-photon detuning, which is the same for both transitions and defined as the difference between the transition frequency $\omega_{1,2}$ and the laser frequency $\omega_{1,2}^{(\mathrm{L})}$. If the single-photon detuning is large compared to the Raman couplings, $|\Delta| \gg |\Omega_1|, |\Omega_2|$,}
and the atoms are initially prepared in a superposition of states $\ket{1}$ and $\ket{-1}$ 
then after
adiabatic elimination of the excited state \cite{Brion_2007}, the effective $2 \times 2$
Hamiltonian describing the dynamics in the $(x,y)$ plane for the
$m_F = \pm 1$ ground state manifold reads 
\begin{equation}
	\mathcal{H}_\mathrm{c} = -\frac{\hbar}{4 \Delta} 
	\begin{pmatrix}
		|\Omega_1|^2 & \Omega_1 \Omega_2^*\\ 
		\Omega_1^* \Omega_2 & |\Omega_2|^2
	\end{pmatrix}.
	\label{adiabaticelimination}
\end{equation}
The off-diagonal elements of this Hamiltonian represent the two-photon Raman coupling between the two ground-state sublevels, while the diagonal elements represent the AC Stark shifts of their energies. 
We consider the case when both Raman beams are Laguerre-Gaussian (LG) modes with the same intensity profiles and opposite OAM projections $m$ and $-m$, so the Raman couplings are dependent on spatial coordinates and also $\Omega_1 = \Omega_2^*$. 
As a result, the atom-light coupling Hamiltonian (\ref{adiabaticelimination}) can be written in cylindrical coordinates $\mathbf{r}=(r,\varphi,z)$ the following form:
\begin{equation}\label{eq:2praman}
	\mathcal{H}_\mathrm{c} = \frac{\hbar \Omega}{2} f(r) \begin{pmatrix} 1 &  e^{-2im\varphi} \\ e^{2im\varphi} & 1 \end{pmatrix},
\end{equation}
where the radial profile is defined 
by the LG mode  
\begin{equation}\label{eq:lg_dist}
	f(r) = e^{m}\left(\frac{r}{r_0}\right)^{2m} e^{-m(r/r_0)^2},
\end{equation}
parametrized in such a way that the peak intensity is located at $r=r_0$ with the peak value $f(r_0)=1$. Intensities of Raman beams and the detuning $\Delta$ are now incorporated into a single parameter, the effective two-photon Raman coupling $\Omega$. This parameter can be either positive or negative depending on the sign of $\Delta$. For red detuning $\Delta > 0$ and $\Omega < 0$, for blue detuning $\Delta < 0$ and $\Omega>0$.

Two eigenvalues of the $2\times 2$ matrix in Eq.~(\ref{eq:2praman}) are $\lambda_1=0$ and $\lambda_2=2$.
This leads to an important difference in the behavior of the system for the cases of red- and blue-detuned Raman couplings. If $\Delta<0$, then the ground state energy of an atom corresponds to $\lambda_1$ and is not shifted by the Raman coupling. On the contrary, in the case of red detuning $\Delta>0$, the ground state corresponds to $\lambda_2$, which means that coupling beams will produce a spatially inhomogeneous ground-state energy shift $-\hbar\Omega f(r)$ and, therefore, generate an additional trapping effect for the BEC.
This additional trapping can strongly modify density distribution of the condensate.
In the numerical results of the next section, we will see the implications of this effect on the ground states of the BEC.  

For an atom trapped in the harmonic potential, the single-particle Hamiltonian 
takes the following form:
\begin{equation}\label{eq:ham0}
	\mathcal{H} = \frac{\hbar^2\nabla^2}{2M}\mathbb{I}_2 + \frac{M}{2} \left(\omega_r^2 r^2 + \omega_z^2 z^2  \right) \mathbb{I}_2 + \mathcal{H}_\mathrm{c},
\end{equation}
consisting of (spin-independent) kinetic energy and trap potential, as well as the Raman-coupling term $\mathcal{H}_\mathrm{c}$, which couples the two pseudospin components of the condensate wave function
\begin{equation}
	\Psi = \Psi_\mathrm{a} \ket{-1} + \Psi_\mathrm{b} \ket{1} \equiv \begin{pmatrix}
		\Psi_\mathrm{a}\\
		\Psi_\mathrm{b}
	\end{pmatrix}.
\end{equation}

To simplify the analysis, for the rest of the present work we adopt the harmonic units: $r_{\mathrm{osc}} = \sqrt{\hbar/M\omega_r}$ as the unit of length and $e_{\mathrm{osc}} = \hbar\omega_r$ as the unit of energy. 
For the characteristic frequencies of the trap, we also assume $\omega_z \gg \omega_r$, which allows
us to consider the $z$ dimension as frozen and treat the system
as two-dimensional, similarly to the previous studies \cite{PhysRevA.102.063328,PhysRevLett.122.110402}. The single-particle Hamiltonian then takes the following simplified form
\begin{equation}
	\mathcal{H} = -\frac{\nabla^2}{2} \mathbb{I}_2 + \frac{r^2}{2} \mathbb{I}_2 + \mathcal{H}_\mathrm{c}.
\end{equation}
The ground state of a weakly interacting spinor condensate is then defined through a minimization of the Gross-Pitaevskii energy functional
\begin{multline}\label{eq:gpe_energy2d}
	E = \int d\mathbf{r} \left[  
	\begin{pmatrix}
		\Psi_\mathrm{a}^* &
		\Psi_\mathrm{b}^*
	\end{pmatrix}
	\mathcal{H} 
	\begin{pmatrix}
		\Psi_\mathrm{a}\\
		\Psi_\mathrm{b}
	\end{pmatrix}  + \frac{g_\mathrm{aa}}2 |\Psi_\mathrm{a}|^4  \right. \\ \left. + \frac{g_\mathrm{bb}}2 |\Psi_\mathrm{b}|^4  + g_\mathrm{ab} |\Psi_\mathrm{a}|^2 |\Psi_\mathrm{b}|^2
	\right].
\end{multline}
where $g_{ij}$ are (two-dimensional) nonlinear interaction constants between atoms in states $i$ and $j$, which in our dimensionless units are defined as $g_{ij}=a_{ij}\sqrt{8\pi\omega_z}$, with the s-wave scattering length $a_{ij}$ expressed in the units of $r_{\mathrm{osc}}$ and the frequency $\omega_z$ expressed in the units of $\omega_r$.  
We also fix the wave function normalization condition as
\begin{equation}
\int d\mathbf{r} \left(|\Psi_\mathrm{a}|^2 + |\Psi_\mathrm{b}|^2\right) = N,
\end{equation}
where $N$ is the total number of atoms in the condensate. 
Additionally, we assume the interactions to be spin-symmetric $g_\mathrm{aa}=g_\mathrm{bb}=g$ and also limit the present study to the situations when interactions are repulsive and two components of the condensate are fully miscible, i.e., $g>g_\mathrm{ab}>0$.

\section{Quantum phases}

\begin{figure}[tbp]
	\centering
	\includegraphics[width=0.95\linewidth]{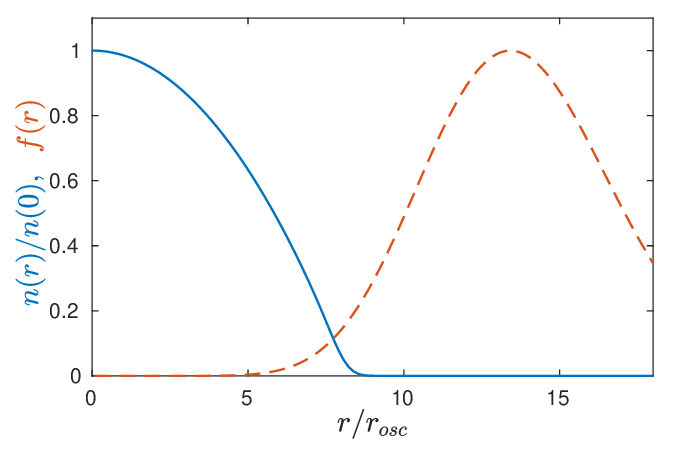}
	\caption{The radial density distribution of the trapped condensate (blue solid line) is shown together with the radial intensity distribution of the Raman coupling (red dashed line).}
	\label{fig:profile}
\end{figure}

Before proceeding with the numerical results and analyzing the obtained ground-state solutions, we need to specify the values for the physical parameters in the system. We consider a system of $N=2\times10^5$ atoms of  $^{23}$Na in a harmonic trap with frequencies $\omega_r=2\pi \times 200$Hz, $\omega_z=2\pi\times 800$Hz, which are typical for the existing cold-atom experiments. This sets the oscillator length to $r_{\mathrm{osc}} = 1.48\,\mathrm{\mu m}$.
The Raman coupling is characterized by Eqs.~(\ref{eq:2praman}) and (\ref{eq:lg_dist}) with  $r_0=20\, \mathrm{\mu m} \approx 13.5\, r_{\mathrm{osc}}$ and $m=5$.    
It is worth noting that many
previous studies have addressed similar systems but with lower orbital angular momentum of LG beams $m = 1$ or $2$. Application of higher-order LG modes is the main difference in the physical setup considered here. 
\revision{While \( m = 5 \) may not appear to be a particularly high OAM value for the LG beam, we also explored higher values of \( m \) and observed that the results remained qualitatively same as those obtained for \( m = 5 \). Nonetheless, a systematic investigation into how the ground state phases depend on the value of \( m \) presents an intriguing avenue for future research.}
It is also important to note that with this choice of parameters, the density distribution of the trapped BEC is concentrated inside the central hole of the LG beams (see Fig.~\ref{fig:profile}).

\begin{figure}[tbp]
	\centering
	\includegraphics[width=0.95\linewidth]{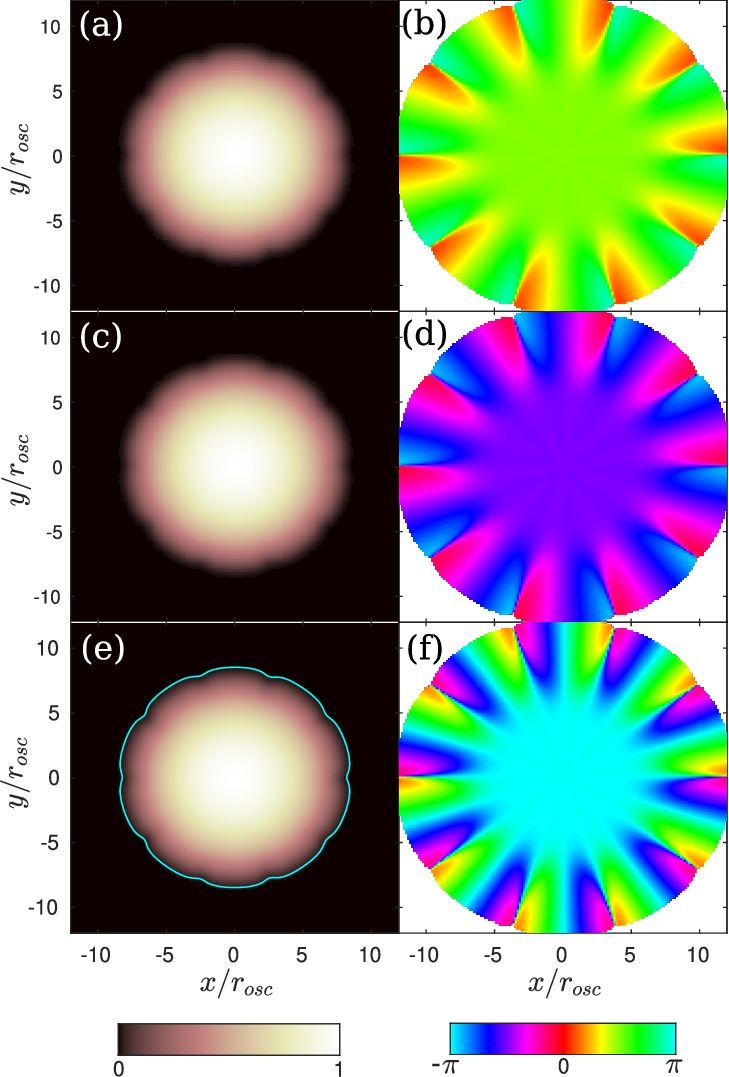}
	\caption{The ground-state wave function for the coupling strength $\Omega=14$. The top row shows (a) the density $|\Psi_\mathrm{a}|^2$ and (b) the phase $\arg\Psi_\mathrm{a}$ of the first component $\Psi_\mathrm{a}$; the second row shows the same for the second component; the third row shows (e) the total density $|\Psi_\mathrm{a}|^2+|\Psi_\mathrm{b}|^2$ and (f) the relative phase $\arg\Psi_\mathrm{a}-\arg\Psi_\mathrm{b}$. The densities are shown in the units of maximum density. \revision{The light blue line on the panel (e) shows the isodensity contour at $1\%$ of the maximum density.}}
	\label{fig:st-phase}
\end{figure}

The Raman coupling strength $\Omega$ remains as a tunable parameter.
We study the dependence of the ground state of the system depending on this parameter and analyze separately the cases of $\Omega > 0$ (blue detuning of Raman transitions) and $\Omega<0$ (red detuning). We begin with $\Omega>0$ as the most interesting case for our study. For the small positive values of $\Omega$, we observe the ground state with the density distribution mostly following the harmonic trap shape and mostly homogeneous phase. When the coupling grows, there are periodic density and phase modulations appearing in both components of the wave function (see Fig.~\ref{fig:st-phase}). 
\revision{The density modulations are relatively weak and can be seen only near the condensate boundary, where the coupling is strongest. To highlight this behavior, we show in Fig.~\ref{fig:st-phase}(e) the isodensity contour at $1\%$ of the maximum density. Deviation of this line from the circular shape reflects the presence of density modulations.}
These states of the system form a so-called ``stripe'' phase, which was previously analyzed in harmonically-trapped systems with SOAM coupling \cite{PhysRevA.102.063328,Chiu_2020,PhysRevResearch.2.033152}. An important property of this phase is its time-reversal symmetry, i.e., $\Psi_\mathrm{a} = \Psi_\mathrm{b}^*$, which can be clearly seen in the density and phase distributions and is also a characteristic symmetry of the system's Hamiltonian \cite{PhysRevA.102.063328}.  

\begin{figure}[tbp]
	\centering
	\includegraphics[width=0.95\linewidth]{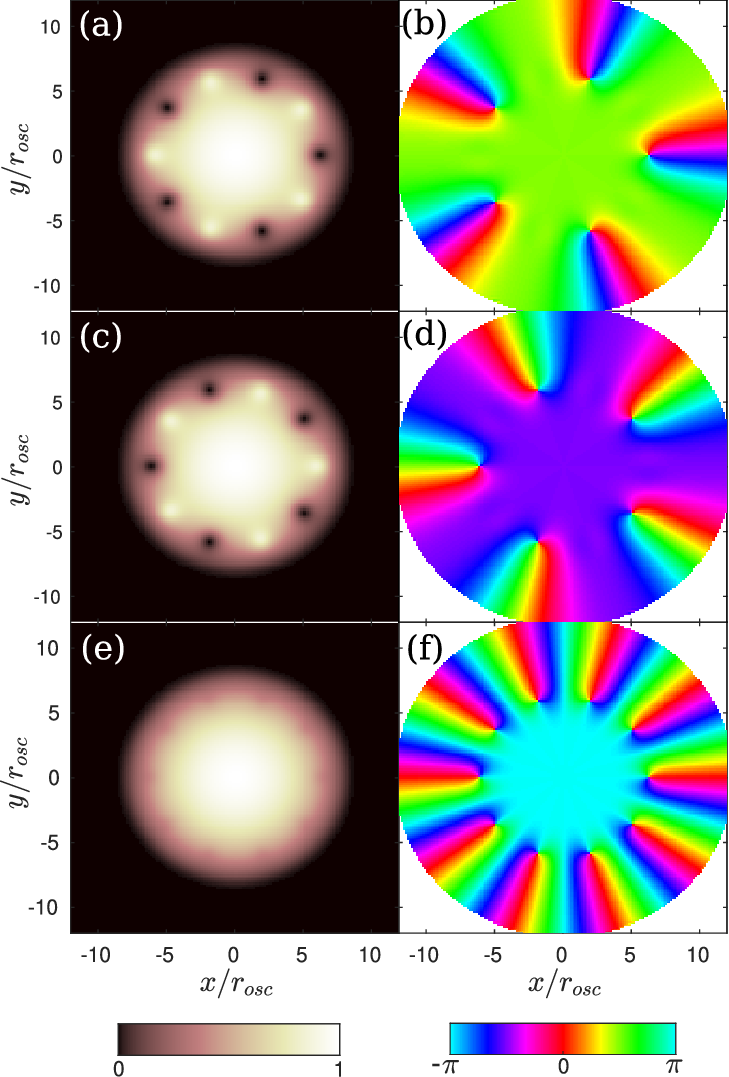}
	\caption{Same as Fig.~\ref{fig:st-phase} but for $\Omega = 36$, which correspond to the vortex-necklace phase.}
	\label{fig:ring-phase}
\end{figure}

When the value of the Raman coupling strength $\Omega$ exceeds a certain critical value $\Omega_\mathrm{cr} \approx 21$, the structure of the ground state changes significantly.  
The ground state now contains $m$ vortices in each spin component (see Fig.~\ref{fig:ring-phase}). All vortices are located at the same distance from the trap center and form a necklace-like pattern. These vortices are known as coreless or polar vortices. Such a vortex is characterized by a phase singularity and condensate density suppression in one component, correlated with the density peak in another, such that the total particle density is almost not perturbed. As the coupling strength increases further, the radius of the vortex necklace decreases.
This phase of the system cannot be directly identified with any of the vortex phases previously observed in SOAM-coupled systems \cite{PhysRevA.92.033615,PhysRevResearch.2.033152,PhysRevA.102.013316}.
One possible interpretation of the described ground-state structure is as follows: due to a strong inhomogeneity of Raman coupling, the system forms two spatially separated regions of two different quantum phases. In the central region inside the vortex necklace, the system remains in the stripe phase. In the outer region we see a ``single-momentum'' phase, which is characterized by the $2m$ winding of the relative phase \cite{PhysRevA.105.023320}. The vortex necklace is then considered as a domain wall between the two regions. In the next section we will use this general understanding to build an analytical model of the vortex-necklace phase.

\begin{figure}[tbp]
	\centering
	\includegraphics[width=\linewidth]{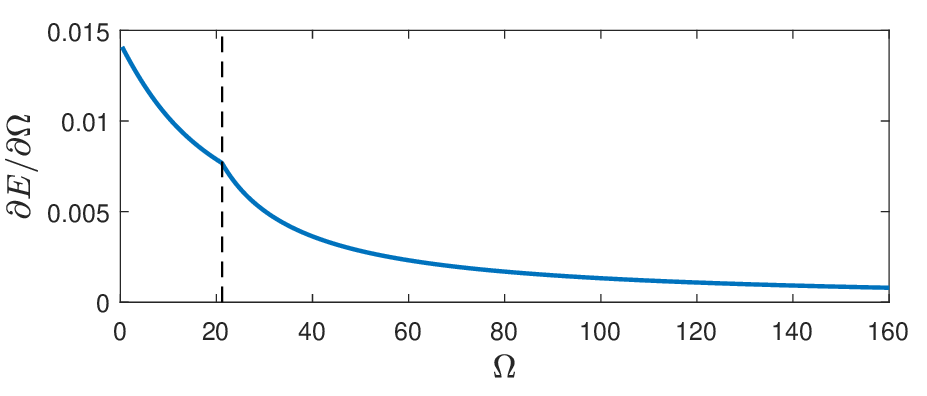}
	\includegraphics[width=\linewidth]{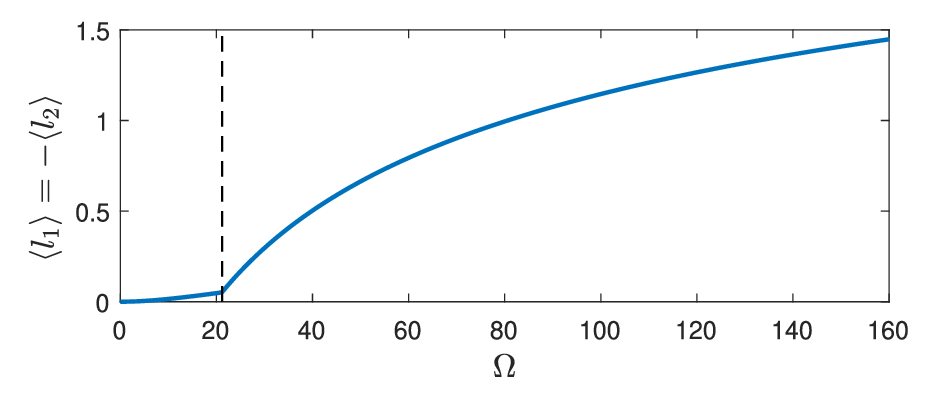}
	\caption{Phase diagram of the system depending on the Raman coupling strength $\Omega$. Top panel shows the derivative of the ground state energy $\partial E/\partial \Omega$. Bottom panel shows the expectation value of the angular momentum of each of the wave function components. The phase transition is observed at $\Omega_{cr} \approx 21$.}
	\label{fig:blue_pd}
\end{figure}

The transition between the two phases described above is a second-order phase transition, which is clearly evident from the behavior of energy derivative $\partial E/\partial\Omega$ (see Fig.~\ref{fig:blue_pd}). Other characteristic physical quantities show a sharp change in their behavior as well. For example, the angular momentum of each condensate component remains close to zero in the stripe phase and starts to grow rapidly in the vortex-necklace phase. 
\revision{From the structure of the wave function in the vortex-necklace phase shown in Fig.~\ref{fig:ring-phase}, it is obvious that $\Psi_\mathrm{a} \neq \Psi_\mathrm{b}^*$, i.e., it is not time-reversal symmetric. We therefore conclude that the observed phase transition is associated with the spontaneous breaking of the time-reversal symmetry.} 

\begin{figure}[tbp]
	\centering
	\includegraphics[width=\linewidth]{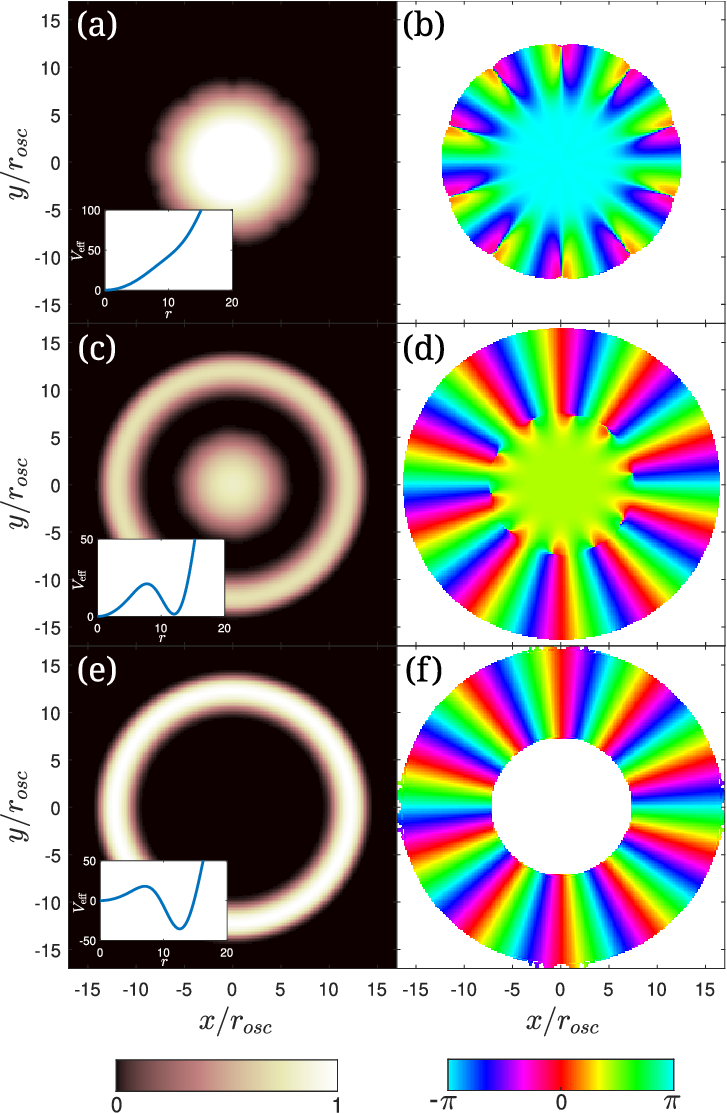}
	\caption{Possible states of the condensate for the red-detuned Raman coupling. Left column shows total particle density $|\Psi_\mathrm{a}|^2+|\Psi_\mathrm{b}|^2$ in the units of maximal density, right column shows relative phase $\arg\Psi_\mathrm{a}-\arg\Psi_\mathrm{b}$. Panels (a), (b) correspond to $|\Omega| = 16$, (c), (d) correspond to $|\Omega| = 80$, (e), (f) correspond to $|\Omega| = 120$.  Insets in panels (a), (c), (e) show the effective trapping potential $V_{\mathrm{eff}}(r) = r^2/2 - \Omega f(r)$.}
	\label{fig:red-states}
\end{figure}

If Raman beams are red-detuned from the single-photon transition, i.e., $\Delta > 0$, we observe a significantly different picture. At small values of Raman coupling we again obtain a familiar stripe phase, similarly to the blue-detuned case. However, as the coupling strength grows, a new behavior emerges. An additional ring-shaped condensate appears, which is located at the region of highest intensity of Raman beams and contain a $2m$ winding of the relative phase. As the coupling grows further, the internal harmonically trapped condensate disappears completely as all atoms redistribute to this outer ring.
Examples of ground states observed with different values of $\Omega$ are shown in Fig.~\ref{fig:red-states}.

\begin{figure}[tbp]
	\centering
	\includegraphics[width=\linewidth]{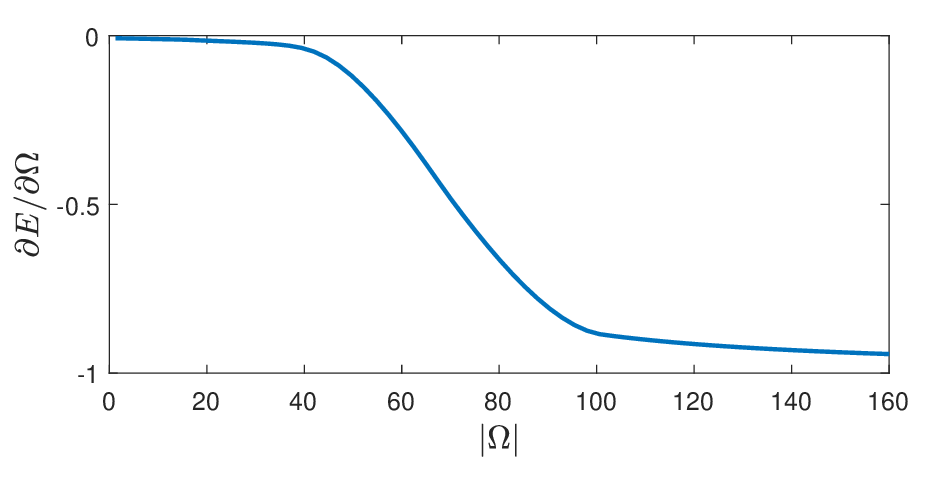}
	\caption{Derivative of the ground state energy with respect to the Raman coupling strength $\partial E/\partial \Omega$ in case of red detuning of the Raman transitions. There are no visible signatures of a phase transition.}
	\label{fig:red_pd}
\end{figure}

To understand this behavior we recall the argument from the previous section: for red detuning the coupling Hamiltonian introduces an additional effective trapping for the condensate. This can be clearly seen comparing the particle density distributions and the shape of the effective potential $V_{\mathrm{eff}}(r) = r^2/2 - \Omega f(r)$ shown in Fig.~\ref{fig:red-states} for different values of $\Omega$. 
It therefore becomes clear that transformations of the ground state observed in the red-detuned case are the result of the effective trap deformation, and are difficult to unambiguously attribute to any phase transitions. In particular, we do not observe any discernible irregularities in the behavior of energy derivative (see Fig.~\ref{fig:red_pd}). It should be noted that our result does not generally exclude the possibility of phase transitions in such a system. Moreover, a previous study \cite{PhysRevA.102.013316} finds real phase transitions and quantum vortices accompanying similar ground-state transformations. Here, however, we demonstrate the possibility of obtaining complex transformations of the ground state without a clear phase transition and without any traceable quantum vortices.

To conclude, we emphasize that the drastic difference between red- and blue-detuned cases in our system is the result of a high inhomogeneity of the Raman coupling, which is a direct consequence of our setup with high-order LG modes. This could be the reason why such effects were not identified in the previous studies with lower-OAM couplings.

\section{Analysis of the mixed phase and the domain wall}

We now consider in more detail the observed states of the system, which exhibit the peculiar necklace of quantum vortices. As already mentioned above, we propose to describe this state not as a single quantum phase, but rather as a state where two phases are spatially separated by a domain wall. A similar phenomenon was previously described in Ref.~\cite{Yao2022}. 

Our numerical results of the previous section show that all the vortices forming the necklace are equally spaced along the ring. Also, vortices in each of the two components are always arranged in alternating order. 
The domain wall radius $R$ remains the only property of the domain wall, which depends on the coupling strength. Therefore, analysis of this dependence will be the main objective of the present section.

To estimate the domain wall radius we have to find a way to analytically minimize the energy functional (\ref{eq:gpe_energy2d}). To this end, we apply several additional approximations to the system. We first introduce the particle density averaged over the angular coordinate as:
\begin{equation}\label{eq:r_profile}
\rho(r) = \frac{1}{2\pi}\int\limits_0^{2\pi} d\phi  \left( |\Psi_\mathrm{a}|^2 + |\Psi_\mathrm{b}|^2 \right),
\end{equation}
and assume that this averaged radial density profile $\rho(r)$ does not depend on the Raman coupling strength $\Omega$. In the limiting case $\Omega=0$, the wave function becomes independent of azimuthal angle, and, therefore, $\Psi_\mathrm{a}=\Psi_\mathrm{b}=\sqrt{\rho}/2$. In the general case of non-zero $\Omega$, we 
rewrite the wave function as follows:
\begin{equation}\label{eq:ang_fun}
\Psi_\mathrm{a,b}(r,\varphi) = \sqrt{2 \pi \rho(r)}\, \psi_\mathrm{a,b}(r;\varphi)
\end{equation}
so that the angular integral
\begin{equation}\label{eq:psi_norm}
\int\limits_0^{2\pi} d\varphi \left( |\psi_\mathrm{a}|^2 + |\psi_\mathrm{b}|^2 \right) = 1
\end{equation}
is independent of the radial coordinate $r$. We also remind the normalization condition for the function $\rho(r)$
\begin{equation}
2\pi \int\limits_0^{\infty} r\, dr\, \rho(r) = N.
\end{equation}
The wave function representation based on Eqs.~(\ref{eq:r_profile})--(\ref{eq:psi_norm}) is similar to a well-established ``pseudospin'' representation \cite{doi:10.1142/S0217979205029602}, commonly used to analyze vortices in binary condensates. The essential difference of our approach is the angular integration in Eqs.~(\ref{eq:r_profile}) and (\ref{eq:psi_norm}), which is motivated by the idea to have a density $\rho(r)$ independent on the angular coordinate and the Raman coupling strength. This will become crucial in the next steps below, especially for drawing parallels between our system and the ring condensates.

After inserting the wave function (\ref{eq:ang_fun}) into the energy functional (\ref{eq:gpe_energy2d}) the total energy of the system becomes \begin{multline}\label{eq:gpe_energy2d_insert}
	E = 2\pi \iint r\, dr\, d\varphi
	\left[
	\frac{(\nabla\sqrt{\rho}\psi_\mathrm{a})^2}2 + \frac{(\nabla\sqrt{\rho}\psi_\mathrm{b})^2}2 
	\right. \\ \left. + \frac{\rho\, \omega_r^2 r^2}{2} +
	\frac{\rho\, \Omega f}{2} +
	 \rho \psi_\mathrm{a}^*\psi_\mathrm{b} e^{-2im\varphi} +
	\rho \psi_\mathrm{b}^*\psi_\mathrm{a} e^{2im\varphi} 
	\right. \\  +
	\pi\rho(r) \left( g |\psi_\mathrm{a}|^4 + g |\psi_\mathrm{b}|^4 + 2g_{ab} |\psi_\mathrm{a}|^2 |\psi_\mathrm{b}|^2 \right) 
	\bigg]
	.
\end{multline}
In order to describe the ground state of the system in the mixed phase with the domain wall,
we can split this energy functional into four parts
\begin{equation}\label{eq:energy_4t}
	E = E_0  + E_\mathrm{I} + E_\mathrm{II} + E_{\mathrm{dw}},
\end{equation}
where the first term
\begin{equation}
	E_0 = \pi \int\limits_0^{\infty} r\, dr \left[ \left(\partial_r \sqrt{\rho(r)}\right)^2 + \rho(r) \omega_r^2 r^2 \right]
\end{equation}
originates from the radial inhomogeneity of the total condensate density. It combines the parts of the total energy that are independent of the Raman coupling and the position of the domain wall. The remaining part of the total energy is split into contributions from three spatial regions: the region of the stripe phase (I), the region of the single-momentum phase (II), and the domain wall region in between the two phases. 
The bulk energies of the two phases can be expressed in the following form: 
\begin{equation}
	E_\mathrm{I} = 2\pi \!\int\limits_0^{R}\! r\, dr \rho(r) \varepsilon(r),\quad
	E_\mathrm{II} = 2\pi \!\int\limits_R^{\infty}\! r\, dr \rho(r) \varepsilon(r),
\end{equation}
where the energy density $\varepsilon(r)$ is defined as
\begin{eqnarray}\label{eq:gpe_energy1d}
	\varepsilon(r) &=& \frac{\Omega f}{2} + \int\limits_0^{2\pi} d\varphi \left[ 
	\frac{(\partial_{\varphi}\psi_\mathrm{a})^2 }{2r^2}+
	\frac{(\partial_{\varphi}\psi_\mathrm{b})^2 }{2r^2}
	+ \psi_\mathrm{b}^*\psi_\mathrm{a} e^{2im\varphi}
	\right. \nonumber\\[1mm] 
	&+& \psi_\mathrm{a}^*\psi_\mathrm{b} e^{-2im\varphi}  
	+ 2\pi G_1 \left(|\psi_\mathrm{a}|^2 + |\psi_\mathrm{b}|^2\right)^2
	\nonumber\\[1mm]  
	&+& 
	2\pi G_2 \left(|\psi_\mathrm{a}|^2 - |\psi_\mathrm{b}|^2\right)^2
	\bigg],
\end{eqnarray}
with 
$G_1 = \rho(g+g_\mathrm{ab})/4$,
$G_2 = \rho(g-g_\mathrm{ab})/4$, and we have neglected the terms that include $\partial_r \psi_\mathrm{a,b}$. This is the next approximation we have to make. Namely, we assume that away from the domain wall the wave functions $\psi_\mathrm{a,b}$ very weakly depend on $r$. As a result of this approximation we obtain the  energy functional (\ref{eq:gpe_energy1d}), which is equivalent to the energy of a thin condensate ring of the radius $r$. It was previously analyzed in detail in Ref.~\cite{PhysRevA.105.023320} and we will adopt some of the analytical results obtained there.

The last term in Eq.~(\ref{eq:energy_4t}) is the energy associated with the domain wall $E_\mathrm{dw}(R)$, which is more difficult to estimate than other terms. We consider that this energy mainly originates from particle density deformation around the vortex cores. Using a common analytical model of the near-vortex density distribution \cite{PhysRevLett.116.160402,PhysRevA.97.063615,doi:10.1142/S0217979205029602}, the value of this energy can be estimated as (see Appendix~A for derivation) 
\begin{equation}
	E_\mathrm{dw}(R) = m \pi \left(2-\log 2\right) \rho(R).
\end{equation}

The stationary state of the system, characterized by the energy functional (\ref{eq:energy_4t}), can be defined by the condition $\partial E / \partial R = 0$, which yields the following expression
\begin{equation}\label{eq:phase-trans-cond1}
	2\pi R \rho(R) \varepsilon_\mathrm{I}(R) - 2\pi R \rho(R) \varepsilon_\mathrm{II} (R) + \frac{\partial E_\mathrm{dw}}{\partial R} = 0,
\end{equation}
where $\varepsilon_{\mathrm{I}}$ and $\varepsilon_{\mathrm{II}}$ are two minima of the energy functional (\ref{eq:gpe_energy1d}) which correspond to stripe and single-momentum phase. These energies can be found within the variational approximation as follows (see Appendix~B for more details):
\begin{equation}
	\begin{aligned}
		\varepsilon_{\mathrm{I}}(R) &= G_1 - \frac{\Omega^2 f^2}{8(G_1+m^2/R^2)}, \\
		\varepsilon_{\mathrm{II}}(R) &= G_1 + \frac12 \left(\frac{m^2}{R^2} - \Omega f \right).
	\end{aligned}
\end{equation}
Eq.~(\ref{eq:phase-trans-cond1}) can be further rewritten in the following simpler form:
\begin{equation}\label{eq:phase-trans-cond2}
\varepsilon_\mathrm{I}(R) - \varepsilon_\mathrm{II} (R) - \chi(R) = 0,
\end{equation}
which is similar to a usual condition for a phase transition, i.e. equality of energy densities of the two phases, modified by an additional term
\begin{equation}\label{eq:dw-tension-def}
\chi(R) = - \frac{1}{2\pi R \rho(R)} \frac{\partial E_\mathrm{dw} (R)}{\partial R}.
\end{equation}
This term can be referred to as the domain wall tension. Using the explicit form of the energy $E_\mathrm{dw}$ this quantity can be expressed as follows
\begin{equation}
\chi(R) = -m\left(1-\frac{\log 2}{2} \right) \frac{\rho'(R)}{R \rho(R)}
\end{equation}

\begin{figure}[tbp]
	\centering
	\includegraphics[width=\linewidth]{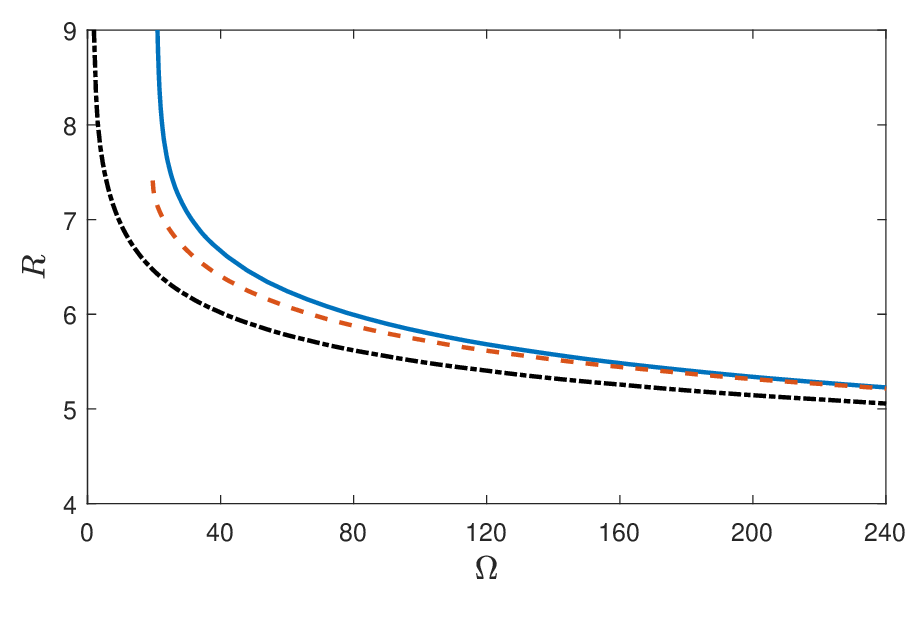}
	\caption{Radius of the vortex necklace (domain wall) as a function of the Raman coupling strength $\Omega$. Blue solid line represents the result extracted from numerically calculated ground state, red dashed line is the solution of Eq.~(\ref{eq:r_ring_cond}), black dash-dotted line is the result from the same equation, but neglecting the domain wall tension ($\chi=0$).}
	\label{fig:r_vortex}
\end{figure}

The solution of Eq.~(\ref{eq:phase-trans-cond2}) yields the radius of the domain wall as a function of $\Omega$. While it cannot be analytically solved for $R$, it can be solved for $\Omega$, resulting in an inverse function $\Omega(R)$, which is found in the following explicit form
\begin{equation}\label{eq:r_ring_cond}
\Omega = \frac{2}{f(R)} \left\{  \frac{m^2}{R^2}+G_1 
-\sqrt{\big[G_1-2\chi\big]\!\left[\frac{m^2}{R^2}+G_1\right]}\right\} , 
\end{equation}
where $G_1$ and $\chi$ are also functions of $R$ through their dependence on the radial density $\rho(R)$. In Fig.~\ref{fig:r_vortex} the result of Eq.~(\ref{eq:r_ring_cond}) is compared to the domain wall radius, found from the full numerical minimization of the Gross-Pitaevskii energy functional (\ref{eq:gpe_energy2d}).

To highlight the importance of the domain wall tension, we also show for comparison the result obtained with $E_\mathrm{dw}=0$ and consequently $\chi=0$ in (\ref{eq:r_ring_cond}). The resulting dependence of the domain wall radius on the coupling strength still qualitatively agrees with the numerical curve but shows a significantly greater discrepancy. Most importantly, such an approximation has completely failed to predict the phase transition in the system. It just shows the radius diverging at $\Omega \rightarrow 0$.

Our model for the domain wall energy works better when the domain wall is closer to the center but becomes less accurate as the radius of the domain wall increases. This is primarily because the size of the vortex cores increases as they shift to lower-density regions of the condensate. Consequently our assumption of having one phase below $R$ and the other one above becomes inaccurate and one would need to introduce a finite thickness of the domain wall between the phases.

It is also important to note that Eq.(\ref{eq:r_ring_cond}) does not predict the domain wall radius diverging at a finite $\Omega$, so technically it does not show the phase transition in the same way as our numerical result. Nevertheless, Eq.~(\ref{eq:r_ring_cond}) has no real roots when the expression under the square root becomes negative, i.e., if $2\chi > G_1$. This physically means that the energy difference between the two phases is smaller then the energy needed to create the domain wall. This transition is observed at $\Omega = 19.6$, which is close to the actual phase transition point $\Omega_\mathrm{c} \approx 21$. 

\section{Conclusions}

This study has examined the quantum phase behavior and ground-state structures of a two-component Bose-Einstein condensate with spin-orbital-angular-momentum coupling (SOAMC) influenced by high orbital angular momentum and pronounced inhomogeneity in the coupling strength. 
Through variation of the Raman coupling strength, we identified possible quantum phases and phase transitions.
Our results reveal that the condensate's response to SOAMC is highly dependent on the detuning of the Raman coupling. For blue-detuned lasers, we observed a distinct quantum phase transition between the stripe and vortex-necklace phases, with the vortex necklace functioning as a boundary separating a central stripe phase from an outer single-momentum phase. This transition underscores the sensitivity of the condensate's structure to variations in coupling strength, as high inhomogeneity encourages the formation of topologically intricate structures. By contrast, red-detuned coupling does not induce such a clear phase separation but rather modifies the condensate's geometry, reorganizing it into a ring-like structure due to changes in the effective trapping potential.

To understand the properties of the vortex necklace, we developed an analytical model estimating the radius of this structure as a function of coupling strength. While the model accurately predicts the domain wall radius for large couplings, it becomes less precise with lower values due to the limitations of assuming the size of vortices to be negligibly small. Nevertheless, this model offers valuable insights into how inhomogeneous SOAMC influences vortex formation and their spatial organization within the condensate.

Overall, this research highlights the complex interplay between orbital angular momentum and spatially inhomogeneous coupling in shaping the phase landscape of SOAMC-driven condensates. Most notably, the observed phase diagram differs significantly from those obtained in previous studies with lower-order LG modes. While the vortex necklace phase can be seen as an extension of previously observed phases with vortices located symmetrically at the same distance from the trap center (see e.g. \cite{PhysRevA.92.033615,PhysRevA.91.053630}), it is not fully clear why the rich variety of other vortex configurations observed with lower OAM is not present in our system. The concept of a mixed quantum phase and the corresponding domain wall is also a unique feature specific to the high-order LG modes and strong inhomogeneity of the Raman coupling. We believe that the model developed in this study enhances our understanding of quantum phase transitions and topological structures in ultracold atomic systems, providing a foundation for further investigations into SOAMC-induced phenomena in condensed matter systems. 
  
\begin{acknowledgments} 
	O.O.P. acknowledges support from the National Research Foundation of Ukraine grant (2023.03/0097) ``Electronic and transport properties of Dirac materials and Josephson junctions''. 
	L.V.Z.'s work is supported by the Centre for the Collective Use of Scientific Equipment  ``Laboratory of High Energy Physics and Astrophysics'' of Taras Shevchenko National University of Kyiv. 
\end{acknowledgments} 

\appendix

\section{Energy of the domain wall}\label{app:vortex_energy}

The energy contribution of the domain wall originates from the particle density deformation around the vortex cores. To calculate this energy we adapt here a common model of vortices in binary condensates used e.g. in Refs. \cite{PhysRevLett.116.160402,PhysRevA.97.063615,doi:10.1142/S0217979205029602}, which compares nicely to our numerical profiles. One key approximation we need to make here is that the vortex core is much smaller then the size of the condensate, so we can model a vortex core as a deformation on top of a uniform background density. Vortices in our system are coreless vortices, which means that density dip in one component is combined with the density peak in another. For the vortex located at the point $\mathbf{r}_0$ densities of both condensate component can be expressed as follows:
\begin{eqnarray}
\rho_\mathrm{a}(\mathbf{r}) = \frac{\rho}{2} \, \frac{|\mathbf{r}-\mathbf{r}_0|^2}{|\mathbf{r}-\mathbf{r}_0|^2+4\xi^2},
\\
\rho_\mathrm{b}(\mathbf{r}) = \rho-\rho_\mathrm{a}(\mathbf{r}) = \frac{\rho}{2}\, \frac{|\mathbf{r}-\mathbf{r}_0|^2+8\xi^2}{|\mathbf{r}-\mathbf{r}_0|^2+4\xi^2},
\end{eqnarray}
where $\rho=\rho(\mathbf{r}_0)$ is the background particle density at the point $\mathbf{r}_0$ which is equally distributed between both components, $\xi$ is the spin healing length, which is defined in our dimensionless units as
\begin{equation}
\xi = \sqrt{\frac{1}{2(g-g_\mathrm{ab})\rho}}.
\end{equation}
The energy contribution of the above density deformation consist of two components: quantum pressure contribution and nonlinear interaction contribution. Both of them can be calculated analytically under an assumption of uniform background density. 
The quantum pressure contribution:
\begin{equation}
E_\mathrm{qp} = \sum_{j=a,b}\int r d\mathbf{r} \frac{\left(\nabla \sqrt{\rho_j}\right)^2}{2} = \frac{\pi \rho}2 (1-\log 2).
\end{equation}
The nonlinear interaction contribution is defined as the difference between the nonlinear interaction energies calculated with vortex profiles $\rho_{a,b}$ and the one calculated with the background profile $\rho$:
\begin{multline}
	E_\mathrm{nl} = \int r d\mathbf{r} \left( \frac{g}{2} \rho_a^2 + \frac{g}{2} \rho_b^2 + g_{ab} \rho_a \rho_b - \frac{g}4 \rho^2 - \frac{g_{ab}}4 \rho^2 \right) \\ 
	= \pi(g-g_{ab})\xi^2\rho^2 = \frac{\pi \rho}2.
\end{multline}
Quite interestingly both components are simply proportional to the background particle density with some constant factor and do not explicitly depend on the other parameters of the system.
The total energy contribution from $2m$ vortices located at the distance $R$ from the center is then:
\begin{equation}
E_\mathrm{dw}(R) = 2m (E_\mathrm{qp} + E_\mathrm{nl}) = m \pi \left(2-\log 2\right) \rho(R),
\end{equation}
where $\rho$ is now the azimuthally symmetric background density profile of the  condensate defined in Eq.~(\ref{eq:r_profile}).
From here we can calculate the tension of the domain wall $\chi$ as it is defined in Eq.~(\ref{eq:dw-tension-def})
\begin{equation}
	\begin{aligned}
		\chi(R) &= \frac{1}{2 \pi R \rho(R)} \frac{\partial E_\mathrm{dw} (R)}{\partial R} \\
		&= m\left(1-\frac{\log 2}{2} \right) \frac{\rho'(R)}{R \rho(R)}.
	\end{aligned}
\end{equation}
We see that this quantity is fully defined by the particle density and its gradient.

\section{Quantum phases of the angular energy density functional}\label{app:ring_energy}

Here we briefly review the main ideas of the variational analysis of the relevant phase transitions in the system, described by the following energy functional
\begin{eqnarray}\label{eq:gpe_energy1d-app}
	\varepsilon = \int\limits_0^{2\pi} d\varphi \left[  
	\begin{pmatrix}
		\psi_\mathrm{a}^* &
		\psi_\mathrm{b}^*
	\end{pmatrix}
	\left( -\frac{\partial_\varphi^2}{2R^2} \mathbb{I}_2 + \mathcal{H}_\mathrm{c}  \right) 
	\begin{pmatrix}
		\psi_\mathrm{a}\\
		\psi_\mathrm{b}
	\end{pmatrix} \right. \nonumber\\ \left. 
	+ 2\pi G_1 \left(|\psi_\mathrm{a}|^2 + |\psi_\mathrm{b}|^2\right)^2
	+ 2\pi G_2 \left(|\psi_\mathrm{a}|^2 - |\psi_\mathrm{b}|^2\right)^2
	\right],
\end{eqnarray}
which is equivalent to Eq.~(\ref{eq:gpe_energy1d}) and describes the thin condensate ring with the radius $R$ and SOAM coupling defined by Eq.~(\ref{eq:2praman}).
Here $\psi_\mathrm{a}$ and $\psi_\mathrm{b}$ are one-dimensional functions of the angular coordinate $\varphi$.
The simplest variational wave function, that is able to capture both the stripe phase and the single-momentum phase, is a two-mode approximation and can be constructed in the following form \cite{PhysRevA.91.063627,PhysRevA.105.023320}: 
\begin{multline}
	\begin{pmatrix}
		\psi_\mathrm{a}\\
		\psi_\mathrm{b}
	\end{pmatrix}
 = \sqrt{\frac{1+\beta}{4\pi}} 
	\begin{bmatrix}
		\cos (\theta/2) e^{-i(m-k)\varphi}\\
		-\sin (\theta/2) e^{i(m+k)\varphi}\\
	\end{bmatrix}
\\ +
\sqrt{\frac{1-\beta}{4\pi}} 
\begin{bmatrix}
	-\sin (\theta/2) e^{-i(m+k)\varphi}\\
	\cos (\theta/2) e^{i(m-k)\varphi}\\
\end{bmatrix},
\end{multline}
where $\beta \in [0,1]$, $\theta \in [0,\pi]$ and $k \in [0,m]$ are variational parameters. Values of these parameters are found by minimization of the energy 
\begin{multline}\label{eq:en-2wave}
	\varepsilon(k,\theta,\beta) = \frac{m^2 + k^2}{2R^2} - \frac{m k}{R^2} \cos\theta  - \frac{\Omega f}2\sin\theta \\ + G_1 + G_1\frac{1-\beta^2}{2}\sin^2\theta  + G_2 \beta^2 \cos^2\theta.
\end{multline}
This variational energy is a linear function of $\beta^2$. Therefore the energy minimum is achieved at one of the limit values $\beta=0$ or $\beta=1$, which corresponds to two possible phases of the system.
First, 
stripe phase is defined by the values of variational parameters $\beta=0$, $k=m$, $\sin\theta\approx\Omega/(2G_1+2m^2/R^2)$ and the energy minimum value
\begin{equation}
	\varepsilon_{\mathrm{I}} = G_1 - \frac{\Omega^2 f^2}{8(G_1+m^2/R^2)}.
\end{equation}
Second, 
single-momentum phase is characterized by $\beta=1$, $k=0$, $\theta=\pi/2$. The corresponding energy is 
\begin{equation}
	\varepsilon_{\mathrm{II}} = G_1 + \frac12 \left(\frac{m^2}{R^2} - \Omega f \right).
\end{equation}
The above simplified variational picture does not cover all possible phases in the system described by Eq.~(\ref{eq:gpe_energy1d-app}), but the above two phases correspond to the ones observed in our harmonically trapped condensate. We therefore restrict our analysis to the two above phases only and refer the interested reader to Ref.~\cite{Chiu_2020} for more extended discussion of the variational treatment in a harmonically trapped system and to Ref.~\cite{PhysRevA.105.023320} for a ring system.
	\bibliography{refs}
\end{document}